\documentclass[fleqn,usenatbib]{mnras}

\usepackage{newtxtext,newtxmath}
\DeclareUnicodeCharacter{2212}{-}
\DeclareUnicodeCharacter{2009}{\,}

\usepackage[T1]{fontenc}

\DeclareRobustCommand{\VAN}[3]{#2}
\let\VANthebibliography\thebibliography
\def\thebibliography{\DeclareRobustCommand{\VAN}[3]{##3}\VANthebibliography}

\usepackage{graphicx}	
\usepackage{xcolor}
\usepackage{amsmath}	

\title[The First Quiescent Galaxies in TNG300]{The First Quiescent Galaxies in TNG300}

\author[A. I. Hartley et al.]{Abigail I. Hartley 
$^{1}$, 
Erica J. Nelson$^{1}$, 
Katherine A. Suess$^{2,3}$, 
Alex M. Garcia$^{4}$, 
Minjung Park$^{5}$,  
\newauthor
Lars Hernquist$^{6}$,
Rachel Bezanson 
$^{7}$,
Rebecca Nevin 
$^{8}$ ,
Annalisa Pillepich 
$^{9}$,
Aimee L. Schechter 
$^{1}$, 
\newauthor
Bryan A. Terrazas 
$^{10}$, 
Paul Torrey 
$^{4}$, 
Sarah Wellons 
$^{11}$,
Katherine E. Whitaker 
$^{12, 13}$, 
Christina C. Williams$^{14, 15}$  
\\
$^{1}$ Department for Astrophysical and Planetary Science, University of Colorado, Boulder, CO 80309, USA\\
$^{2}$ Department of Astronomy and Astrophysics, University of California, Santa Cruz, 1156 High Street, Santa Cruz, CA 95064 USA \\
$^{3}$ Kavli Institute for Particle Astrophysics and Cosmology and Department of Physics, Stanford University, Stanford, CA 94305, USA \\
$^{4}$ Department of Astronomy, University of Florida, 211 Bryant Space Sciences Center, Gainesville, FL 32611, USA \\
$^{5}$ Center for Astrophysics $\mid$ Harvard $\&$ Smithsonian, 60 Garden St., Cambridge, MA 02138, USA \\
$^{6}$ Center for Astrophysics, Harvard University, Cambridge, MA 02138, USA\\
$^{7}$ Department of Physics and Astronomy and PITT PACC, University of Pittsburgh, Pittsburgh, PA 15260, USA\\
$^{8}$ Fermi National Accelerator Laboratory, Batavia, IL 60510, USA \\
$^{9}$ Max-Planck-Institut für Astronomie, K'onigstuhl 17, D-69117 Heidelberg, Germany\\
$^{10}$ Center for Astrophysics, Harvard $\&$ Smithsonian, 60 Garden St., Cambridge, MA, 02138, USA\\
$^{11}$ Department of Astronomy, Van Vleck Observatory, Wesleyan University, 96 Foss Hill Drive, Middletown, CT 06459, USA \\
$^{12}$ Department of Astronomy, University of Massachusetts, Amherst, MA 01003, USA\\
$^{13}$ Cosmic Dawn Center (DAWN), Denmark\\
$^{14}$ Steward Observatory, University of Arizona, 933 North Cherry Avenue, Tucson, AZ 85721, USA\\
$^{15}$ NSF's National Optical-Infrared Astronomy Research Laboratory, 950 North Cherry Avenue, Tucson, AZ 85719, USA}

\date{Accepted XXX. Received YYY; in original form ZZZ}

\pubyear{2022}

\begin{document}
\label{firstpage}
\pagerange{\pageref{firstpage}--\pageref{lastpage}}
\maketitle

\begin{abstract}

We identify the first quiescent galaxies in TNG300, the largest volume of the IllustrisTNG cosmological simulation suite, and explore their quenching processes and time evolution to $z=0$. 
We find that the first quiescent galaxies with stellar masses $\mathrm{M}_* > 3 \times 10^{10} \mathrm{M}_\odot$ and specific star formation rates sSFR $< 10^{-11} \mathrm{yr}^{-1}$ emerge at $z\sim 4.2$ in TNG300.
Suppression of star formation in these galaxies begins with a thermal mode of AGN feedback at $z\sim 6$, and a kinetic feedback mode acts in each galaxy by $z\sim 4.7$ to complete the quenching process, which occurs on a time-scale of $\sim 0.35$ Gyr. Surprisingly, we find that 
the majority of these galaxies are not the main progenitors of their $z = 0$ descendants; instead, four of the five galaxies fall into more massive galaxies in subsequent mergers at a range of redshifts $2.5 < z < 0.2$. By $z = 0$, these descendants are the centres of galaxy clusters with average stellar masses of $8 \times 10^{11} \mathrm{M}_\odot$. 
We make predictions for the first quenched galaxies to be located by the \emph{James Webb Space Telescope} (\emph{JWST}).

\end{abstract}

\begin{keywords}
galaxies: high-redshift -- galaxies: evolution -- galaxies: star formation
\end{keywords}



\section{Introduction}
\label{intro}

By the time the Universe was 3 billion years old, half of its massive galaxies had stopped forming new stars \citep[][]{2021Natur.597..485W, 2013ApJ...777...18M}, but the variations in the time-scales and structural factors involved in their quenching processes often evade explanation. These quiescent galaxies appear to be unable to form stars because they lack reservoirs of cold, star-forming gas
\citep{2019ApJ...873L..19B, 2021ApJ...908...54W, 2021Natur.597..485W}. 
When observed at high redshifts, 
they are exceedingly compact \citep{2009Natur.460..717V, 2013ApJ...775..106C, 2015ApJ...808L..29S, 2014ApJ...788...28V}, which in principle allows for more expedient gas accretion onto their supermassive black holes (SMBHs). If a galaxy's central SMBH accretes rapidly enough, an active galactic nucleus (AGN) is formed, which serves as an incredibly luminous and persistent source of electromagnetic radiation \citep{2022MNRAS.513.3768I, 2017NatAs...1E.165H, 2019lgei.confE..74H}.
By preventing further accumulation of cold gas and dust in the central regions of a galaxy \citep{2018ApJ...866...91C, 2019MNRAS.483.4586F}, AGN feedback is proposed to prohibit quiescent galaxies from rejuvenating and developing dense stellar cores later in time (but see also \citet{2022ApJ...940...39W}
).
This feedback from early black holes has thus likely played a pivotal role in the quenching of early and massive galaxies \citep{2017MNRAS.472..949B, 2021MNRAS.500.4004D, 2022arXiv220402205N}, but causal evidence linking AGN feedback to quenching has been nearly impossible to establish observationally. This empirical uncertainty necessitates the use of cosmological simulations in studies of AGN feedback.

In the IllustrisTNG (hereafter TNG) cosmological simulation suite \citep{2018MNRAS.475..624N,  2018MNRAS.475..648P} modeling galaxy formation, AGN feedback regulates black hole accretion and star formation with quasar-heating (thermal) and radio-jet (kinetic) wind modes, which work at different scales to expel cool star-forming gas from their host galaxies \citep{2007MNRAS.380..877S, 2022NatAs...6..488M}. The thermal mode heats surrounding gas within a galaxy, while the kinetic mode serves as a velocity boost to eject the gas.
Both 
feedback modes may work in tandem to create and maintain the quiescence of their host galaxy \citep{2017FrASS...4...42M, 2018MNRAS.477.1336C}. 

Although many studies have been conducted to determine the physical properties and quenching processes of ancient quiescent galaxies 
\citep[e.g.,][]{2015ApJ...813...23V,  2021arXiv211207679P, 2022MNRAS.513.4814Z, 2022arXiv220109068L}, TNG provides a means to address a cosmological enigma yet to be resolved in this field, namely the lives and quenching mechanisms of the very first quiescent galaxies in our Universe. 
The most massive galaxies in the Universe today host maximally old stars \citep[e.g.,][]{2010MNRAS.404.1775T, 2015MNRAS.448.3484M}, suggesting that they are relics of very early star formation and the early truncation thereof. This empirical fact has naturally driven observational searches for the earliest progenitors of these systems, and
examples of these massive, quiescent galaxies have been spectroscopically confirmed as early as $z\sim4$ \citep{2017Natur.544...71G, 2020ApJ...889...93V, 2020ApJ...903...47F}.
However, the intrinsically red colours and high mass-to-light ratios of older stellar populations have rendered direct study of the earliest red-and-dead systems observationally out of reach prior to vastly improved sensitivity, wavelength coverage, and resolution of the \emph{JWST}. 
Investigating the evolution and quenching processes of the first massive quiescent galaxies in the largest volume of the TNG suite provides some theoretical underpinnings for the groundbreaking examinations of quiescent galaxies that observational missions like
\emph{JWST} will produce.

In this letter, we shed light on the physical properties, star formation histories, quenching processes (AGN feedback modes), and evolution of the first quiescent galaxies in the TNG300 simulation. 
The letter is organised as follows. In Section~\ref{methodology}, we give a brief overview of the IllustrisTNG project and simulation utilized in this study. We then describe criteria for selection of the first quiescent galaxies within TNG300. 
In Section~\ref{results}, we set forth our results regarding the first quenched galaxies identified in TNG, as well as critical information about their physical properties and star formation histories (Section~\ref{phys}). We then discuss the sample's evolution to $z=0$
(Section~\ref{mergers}) and explore the quenching mechanisms acting on these galaxies (Section~\ref{quenching}).
Section~\ref{jwst} briefly relays the final conclusions drawn from this work, including its applications to the \emph{JWST} mission.

\section{Methodology}
\label{methodology}


\subsection{The TNG300 simulation}

This study utilizes simulations from the IllustrisTNG project \citep{2018MNRAS.475..624N, 2018MNRAS.475..676S, 2018MNRAS.475..648P,2018MNRAS.477.1206N,2018MNRAS.480.5113M,2019MNRAS.490.3196P,2019MNRAS.490.3234N}. IllustrisTNG is a suite consisting of cosmological magnetohydrodynamical simulations run using the moving-mesh {\fontfamily{pbk}\selectfont AREPO} code \citep{2010MNRAS.401..791S}. 
TNG is an updated version of the model employed in the original Illustris
Project \citep{2013MNRAS.436.3031V,2014MNRAS.444.1518V,2014Natur.509..177V, 2014MNRAS.445..175G} and simulates galaxy formation physics 
with a hybrid multiphase model for quiescent star formation coupled with radiative heating and cooling, solving for
the integrated evolution of cosmic gas, luminous stars, dark matter and SMBHs 
\citep{2003MNRAS.339..289S,2018MNRAS.473.4077P}. 
The simulations include SMBH growth with high-accretion state thermal mode \citep{2005Natur.433..604D,2005MNRAS.361..776S, 2007MNRAS.380..877S} and low-accretion state kinetic wind mode feedback \citep{ 2015MNRAS.452..575S,2017MNRAS.465.3291W}.
The instantaneous SMBH mass accretion rates calculated by TNG reflect the expected Bondi accretion rate of each respective SMBH given its mass and the properties of the surrounding gas.

The TNG suite consists of 3 simulation boxes with varying resolutions and volumes, which allows for both narrow and extensive cosmological studies. 
Because we would like to study rare galaxies, we utilize data selected from the TNG300 simulation \citep{ 2019ComAC...6....2N}, which has has a box size of $\sim$ 300 Mpc$^3$, the largest volume of the TNG suite.
The simulation assumes the \citet{2016A&A...594A..13P} best fit cosmological parameters, namely a dark energy density $\Omega_\Lambda = 0.6911$, baryon density $\Omega_b = 0.0486$, matter density $\Omega_m = 0.3089$, Hubble constant $H_0 = 67.74 \; \text{km} \; \text{s}^{-1}\text{Mpc}^{-1}$, spectral index $n_s = 0.9667$, and  normalization $\sigma_8 = 0.8159$.
We employ TNG300-1, the highest resolution realization of TNG300, which includes $\sim2\times2500^3$ resolution elements. TNG300 evolves dark matter particles with mass of $6 \times 10^7  \mathrm{M}_\odot$ and baryonic elements (stellar particles and gas cells) with mass resolution of $1.1 \times 10^7  \mathrm{M}_\odot$.

TNG employs the {\fontfamily{pbk}\selectfont SUBFIND} algorithm \citep{2001MNRAS.328..726S, 2009MNRAS.399..497D} to locate gravitationally bound structures within the simulation. These structures include subhaloes and dark matter haloes, with associated baryonic components that comprise galaxies within each subhalo. Subhaloes are tracked through snapshots in time using {\fontfamily{pbk}\selectfont SUBLINK} merger trees \citep{2015MNRAS.449...49R}.
A subhalo's descendant is the subhalo with the highest weighted sum of individual particles (gas, stars, and dark matter) shared with the progenitor. These particles are ranked by gravitational binding energy and weighted by (rank)$^{-1}$. A merger takes place when multiple subhaloes share a common descendant, and the main progenitor of this descendant is defined as whichever subhalo has the most massive history \citep{2007MNRAS.375....2D}.


\subsection{Selection}

We select galaxies that are quiescent by requiring specific star formation rate (sSFR), defined as star formation rate normalized by galaxy mass within twice the stellar half-mass radius (INRAD quantities in the TNG catalogues), to fall below $ 10^{-11} \text{yr}^{-1}$ \citep{2000ApJ...536L..77B, 1997ApJ...489..559G}. 
We check that the descendant subhaloes of our selection maintain quiescence down to $z=2$,
in order to exclude temporary low-activity galaxies and ensure that we are locating those that are truly quenched.
Finally, we limit galaxy stellar mass $\mathrm{M}_*$  to $\log(\mathrm{M}_*/\mathrm{M}_\odot) > 10.5$.
This restricts our search to subhaloes with stellar masses greater than 1000 times the baryonic mass resolution of TNG300-1, so that all galaxies are resolved with roughly $10^4$ star particles.


\section{Results and Discussion: The first quiescent central galaxies in TNG300}
\label{results}


The first quiescent galaxies in TNG300 emerge at $z\sim 4.2$, 
roughly $1.5$ Gyr after the Big Bang.
No snapshots in time prior to $z=4.18$ contain galaxies that fit our criteria.
To ensure a robust selection, we experimented with a higher sSFR cut of $10^{-10} \text{yr}^{-1}$, and our five selected galaxies were still the only quiescent candidates at $z=4.18$ (with no new galaxies fitting these criteria at higher redshifts). 
Lowering our mass cut to $10^{10} \mathrm{M}_\odot$ resulted in the emergence of a new quiescent galaxy at the prior snapshot of $z=4.43$, which was found to be a galaxy from our sample that hadn't yet reached its star formation peak. We also checked TNG50, the highest resolution simulation of the TNG suite, for galaxies meeting our original criteria, and found the first match to occur at $z=3.0$ (mostly due to a volume effect).
With a larger volume, we may detect even rarer quiescent sources at higher redshifts. We note that the TNG300 simulation's effective comoving area at $z\approx4.2$ 
is roughly 75\% of that covered by the five HST CANDELS survey fields at the same redshift \citep{2020ApJ...890....7C}. By redshift $z=5$, TNG300’s effective comoving area only makes up $\sim50$\% of that of the CANDELS survey, which may indicate compelling discrepancies between observed and simulated samples of high-redshift quiescent galaxies.

\begin{figure}
	\includegraphics[width=\columnwidth]{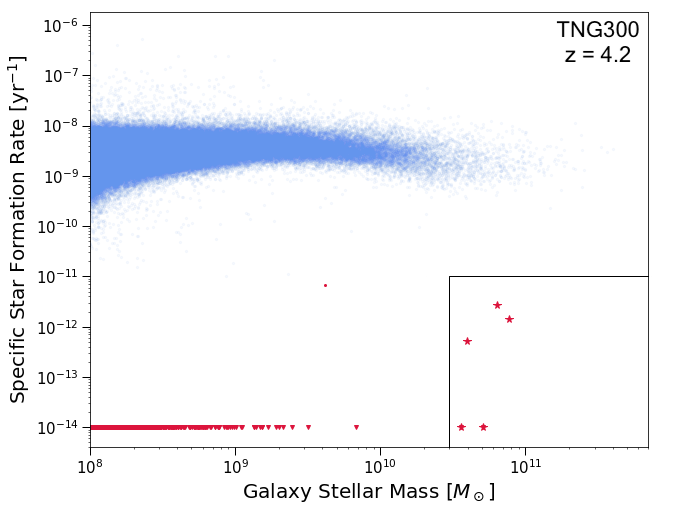}
    \caption{sSFR within twice the stellar half-mass radius of each galaxy with $\mathrm{M}_* > 10^8 \mathrm{M}_\odot$ at $z=4.18$ as a function of total stellar mass in the TNG300 simulation. Galaxies with sSFR $< 10^{-11} \text{yr}^{-1}$ are colored red; all others are colored blue. Galaxies with sSFR$=0$ have been brought up to the band at sSFR $=10^{-14}$ [yr$^{-1}$] in downward carets. 
    Galaxies with stellar masses below $10^8 \mathrm{M}_\odot$ are not resolved and thus excluded from this visualization.
    Our selection criteria of 
    massive quiescent galaxies is boxed; five galaxies satisfy this selection in TNG300 at $z\sim 4.2$.}
    \label{fig:SFRM}
\end{figure}

Fig.~\ref{fig:SFRM} plots sSFR versus galaxy stellar mass for all subhaloes 
with $\mathrm{M}_* > 10^8 \mathrm{M}_\odot$ at $z=4.18$, displaying the clear locus of galaxies at sSFR $\sim 3 \times 10^{-9}$ which make up the star-forming main sequence (SFMS) \citep{2021MNRAS.508..219N}. However, $9\%$ of galaxies fall $> 1$ dex below this main sequence. 
All five galaxies below an sSFR of $10^{-11}$ and above a stellar mass of $3 \times 10^{10} \mathrm{M}_\odot$ have already passed their star formation peaks.

We find the number density of our sample, defined as the number of galaxies that fit our selection criteria at $z\sim 4.2$ divided by the effective volume of the TNG300 simulated box, to be $n = 1.32 \times 10^{-5}~\text{Mpc}^{-3}$.
This closely resembles the observational number densities calculated by \citet{2022arXiv220800986C} reflecting massive quiescent galaxies located by \emph{JWST} at $3 < z < 5$, and by \citet{2018A&A...618A..85S} in their study of massive quiescent galaxies at $3 < z < 4$ identified in the ZFOURGE and 3DHST catalogues \citep{2014ApJS..214...24S, 2016ApJ...830...51S}.
The authors of \citet{2018A&A...618A..85S}  report a number density of $2.0 \pm 0.3 \times 10^{−5} n / \text{Mpc}^{−3}$ for a spectroscopic sample of quiescent galaxies at $3 < z < 4$ 
, and the number density of 
quiescent galaxies in \citet{2022arXiv220800986C}'s robust sub-sample at $3 < z < 4$ is found to be $6.3^{+3.8}_{−2.5} \times 10^{−5} n / \text{Mpc}^{-3}$.
\citet{2020ApJ...889...93V} show that the number densities of these quiescent galaxies at $3 < z < 4$ are consistent between simulations and observations.
Furthermore, the authors of \citet{2022arXiv220800986C} report a number density of $2.3^{+3.1}_{−1.5} \times 10^{−5} n / \text{Mpc}^{−3}$ for massive quiescent galaxies located by \emph{JWST} at $4 < z < 5$.
However, these observed number densities of high-redshift quiescent galaxies are dependent upon measurement choices, such as definitions of quiescence and SFR time-scale, and sample selection functions, which impact their comparison to theoretical investigations \citep{2021MNRAS.506.4760D}. The number densities presented in early \emph{JWST} studies also reflect preliminary observations.
It can be difficult to distinguish dusty star-forming galaxies and quiescent galaxies from photometry alone, so further spectroscopic observations are needed to confirm high-redshift ($z\sim4$) quiescent galaxy candidates.


\subsection{Physical Properties and Star Formation Histories}
\label{phys}

\begin{figure}
	\includegraphics[width=\columnwidth, ]{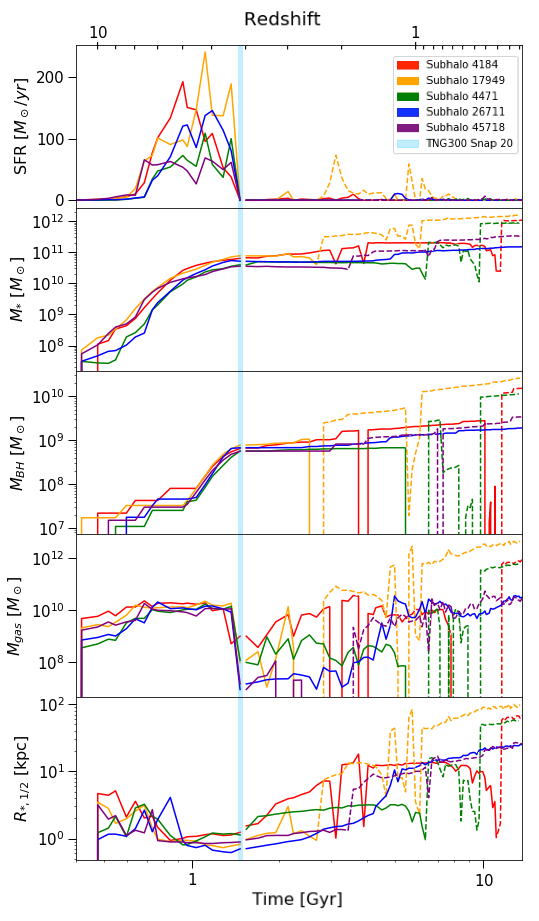}
    \caption{For each galaxy in our sample, 
    this plot shows the time evolution of star formation rate (SFR), stellar mass ($\mathrm{M}_*$), SMBH mass ($\mathrm{M}_{\rm BH}$), gas mass ($\mathrm{M}_{\rm gas}$), and stellar half-mass radius ($R_{*, 1/2}$).
    All quantities are restricted to cells within twice the stellar half mass radius of each galaxy.
    An age of 1.47 Gyr is marked in light blue on each plot to indicate the time of study (Snapshot 20 of TNG300, or $z=4.18$). After galaxies in the sample merge into more massive subhaloes, they are plotted with dashed lines.}
    
    \label{fig:5SFH}
\end{figure}

Fig.~\ref{fig:5SFH} plots 
the time evolution of the star formation rates of our selected galaxies,
their stellar masses, black hole masses, gas masses, and stellar half-mass radii. 
These plots include all mergers experienced by our sample galaxies since $z=4.18$, with a dashed line indicating that the galaxy merged into a more massive halo and is no longer its descendant's main progenitor. 
All galaxies are first identified as quiescent at an age of 1.47 Gyr, marked in blue with a vertical line on each panel of the figure. This line illustrates that our galaxies of interest have indeed passed their epochs of peak star formation.

Our sample displays star formation rates that drop off and approach zero soon before an age of $\sim 1.47$ Gyr, with star formation history shapes defined by early peaks with compact widths.
The average sSFR of this sample at $z=4.18$ is $9.22 \times 10^{-13} \text{yr}^{-1}$, which falls a few orders of magnitude below the sSFR associated with the SFMS at this redshift. 
We find that this sample experiences an expeditious quenching time-scale of $\sim 0.35$ Gyr. The duration of quenching is defined as the time period during which the sSFR drops continuously from its peak value to sSFR $= 1/[20 t_\text{H}(z)]$, where $t_\text{H}(z)$ is the Hubble time at each redshift.
This quenching is reflected as a decrease in the rate of stellar mass growth preceded by a rapid rise in black hole mass for each galaxy. The mass growth of these galaxies' SMBHs thus seems to be an integral driving factor in their quenching.

We find the average stellar mass contained within twice our sample's respective stellar half-mass radii at an age of $1.47 \text{Gyr}$ to be $5.39 \times 10^{10} \mathrm{M}_\odot$. In comparison, the average gas mass within the same radius for each galaxy is $1.82 \times 10^8 \mathrm{M}_\odot$. 
We calculate the mean gas fraction of this sample over the span of its quenching era, $\sim 1.11$ Gyr to $\sim 1.47$ Gyr 
($z=5.23$ to $4.18$), in order to gauge the rate at which
gas depletion occurs within these galaxies. This fraction drops from $0.53$ at $z=5.23$ to $0.25$ at $z=4.43$, and finally to $3.39 \times 10^{-3}$ at $z=4.18$. This suggests that these quiescent galaxies very rapidly lose their gas while quenching, as expected for a sample with such low specific star formation rates \citep[][]{2019ApJ...873L..19B, 2021ApJ...908...54W, 2021ApJ...922L..30W}. These galaxies display an average gas-phase metallicity of $Z = 0.21 Z_\odot$.

The stellar half-mass radii of our sample range from 0.7 to 1.3 kpc when they are first identified as quiescent, which is standard for massive galaxies at $z=4.18$ in TNG300; the average stellar half-mass radius for galaxies with $\mathrm{M}_* > 3 \times 10^{10} \mathrm{M}_\odot$ in the snapshot is a compact 1.29 kpc. Our five selected galaxies also display distinctly disk-like morphologies.



Utilizing stellar formation times from each star particle within each sample galaxy, we find that the oldest star within the first massive quiescent galaxies in TNG300 formed at a high redshift of $z=15.1$.
On average, however, the stars composing these galaxies were born at $z\approx5.49$, which corresponds to a  lookback time of $\sim 12.7$ Gyr. 
With an average stellar age of $\sim 0.5$ Gyr, this sample is consistent with observed post-starburst galaxies \citep{2022ApJ...926...89S}; the first quiescent galaxies in TNG300 quenched quickly.

To explore how these objects would be detected in real surveys, we provide magnitude estimates in H160 and F444W, typical detection bands for the \emph{Hubble Space Telescope} and \emph{JWST}, respectively. We find the average apparent AB magnitude of our sample to be 24.6 in the U rest frame band, which is comparable to H160, and 23.7 in the i rest frame band,  which is similar to F444W observed for $z\sim4$ sources.
These magnitudes indicate that we have located a strong sample of simulated candidates to obtain \emph{JWST} NIRSpec spectra for in the real Universe. The magnitudes also corroborate those reported in \citet{2023arXiv230111413C}.
However, our galaxies do not fall within the standard UVJ cut for quiescent sources, which indicates ambiguity in their detectability in a potential observational survey.
This limitation highlights the inherent difficulties of both locating the first quiescent galaxies in the Universe, and of thoroughly comparing astrophysical observations and theoretical predictions without a proper forward modeling approach of simulated data. 
The TNG300 simulation does not account for extinction from dust, which may also reduce the accuracy of magnitude approximations. Quiescent galaxies generally have limited reserves of dust, but we should still expect a potential disparity of up to half a mag between our above estimates of apparent visual magnitude and the brightnesses of similar galaxies located in our Universe.

\subsection{Evolution to $z=0$}
\label{mergers}

\begin{figure*}
	\includegraphics[width=\textwidth]{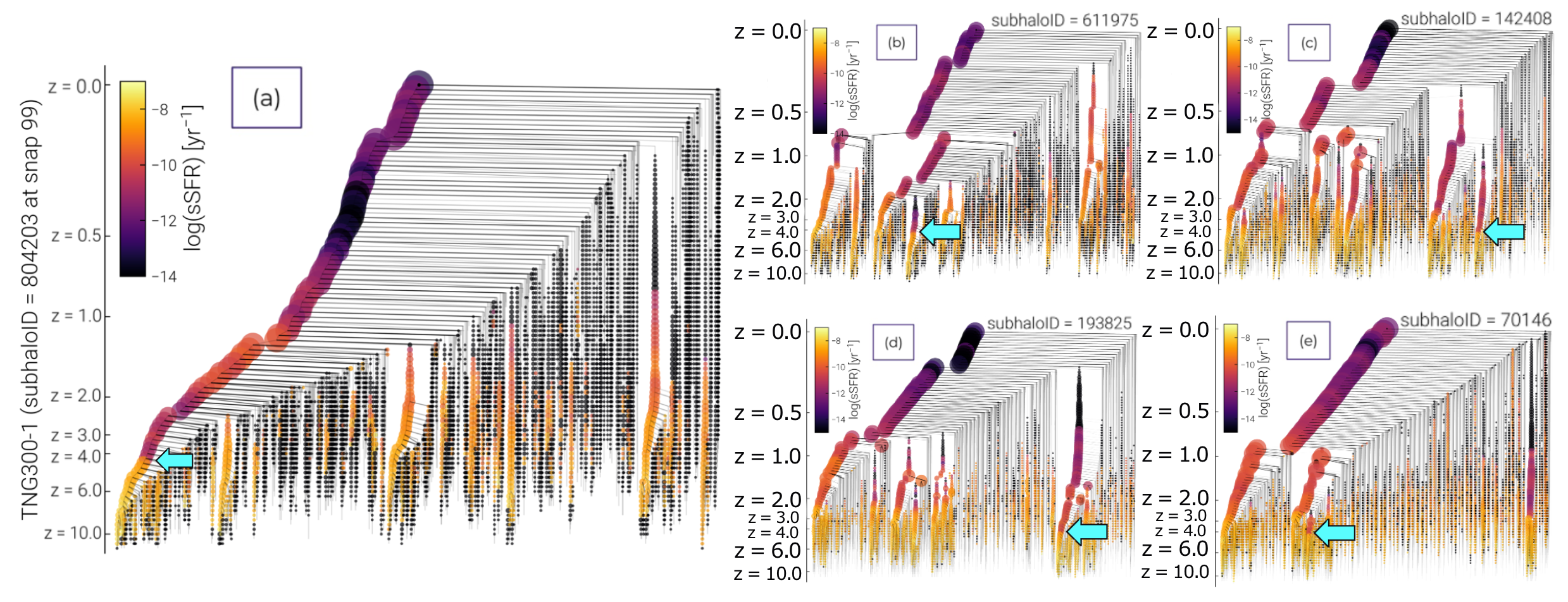}
    \caption{Complete merger trees for (a) the only galaxy in the sample to lie on the main progenitor branch of its $z=0$ descendant and (b) through (e) the remaining galaxies in the sample. Cyan arrows indicate where our sample galaxies lie on these trees. TNG generates these diagrams by tracking all of the galaxies that have merged into the final galaxy at the top of the tree (at redshift $z=0$). Galaxies are color coded according to logarithmic sSFR, and subhalo sizes are determined by stellar mass.}
    \label{fig:trees}
\end{figure*}

Contrary to initial expectations that the first quiescent galaxies in TNG would persist as the most central and massive galaxies of their respective haloes, we find that only one of our five selected galaxies lies on the main progenitor branch of its $z=0$ descendant. The remaining four merge into other, more massive galaxies. 
This can be seen in Fig.~\ref{fig:trees}, which juxtaposes complete merger tree diagrams generated by TNG for the evolutions of: (a) the only galaxy in the sample to lie on the main progenitor branch of its $z=0$ descendant, and (b) through (e), 
the remaining galaxies in the sample which are far removed from their main progenitor branches.

The fact that our selected galaxies in the bottom four panels
do not lie on this branch of the most massive galaxies leading up to their final $z=0$ subhaloes, is a surprise. \citet{2016MNRAS.456.1030W}, for example, found that a mere 15 per cent of their selected massive compact quiescent galaxies in Illustris were consumed in mergers with more massive galaxies on their evolutionary paths from $z=2$ to $z=0$. The majority of galaxies in their sample instead either acquired ex situ envelopes to become the cores of more massive descendants, or survived undisturbed.
\citet{2009ApJ...697.1290B} further describe a model wherein compact high-redshift galaxies comprise the centres of normal nearby ellipticals and grow via minor mergers, which predicts that these galaxies serve as the progenitors of massive elliptical galaxies in the local universe.


In order to explore the large-scale structural transformation of our sample of galaxies, we calculate their physical properties at redshift $z=0$ and determine what they have become.
At $z=0$, the only galaxy in our sample to lie on the main progenitor branch of its $z=0$ descendant sports a stellar half-mass radius of $\sim 24.5$ kpc. The remaining galaxies in our sample have merged into larger galaxy clusters whose centrals have stellar half-mass radii ranging from $25$ to $97$ kpc.  
These radial measurements extend deep into the haloes of the clusters, and would thus make for poor comparisons to observational data.

After a galaxy in the sample merges into another more massive object, it is plotted with a dotted line in Fig.~\ref{fig:5SFH}, which illustrates that a significant amount of this radial increase is due to major mergers into larger galaxy clusters.
We find that the average stellar mass of our sample has increased to $7.94 \times 10^{11} \mathrm{M}_\odot$, and the associated average gas mass has increased to $1.13 \times 10^{12} \mathrm{M}_\odot$. 
Our results suggest that the first quiescent galaxies become brightest cluster galaxies (BCGs) in the local Universe at $z=0$. Though the majority of our sample of galaxies are not the main progenitors of BCGs, they are likely the most massive galaxies contributing to the centres of these clusters.

\subsection{Quenching Processes: SMBH accretion and AGN feedback}
\label{quenching}


At $z\sim 4.2$, the SMBHs within our sample have just experienced rapid mass growth, as evidenced by the middle panel of Fig.~\ref{fig:5SFH}. 
We quantify this accretion by calculating the average black hole growth rate between snapshots. From $z=4.66$ to $z=4.18$, a span of $0.18$Gyr, our sample's central black holes grow by an average of $1.35 \times 10^9 \mathrm{M}_\odot / \text{Gyr}$. We find the average black hole growth rate from $z=4.18$ to $z=0$, a span of $12.34$Gyr, to have declined to stabilize at $4.92 \times 10^8 \mathrm{M}_\odot / \text{Gyr}$, a phenomenon which may be explained by the regulatory nature of AGN feedback in relation to black hole mass.

Our sample's SMBHs exhibit thermal AGN feedback during the entire quenching era of these galaxies, $\sim 1.11$ Gyr to $\sim 1.47$ Gyr after the Big Bang. 
At $1.11 \text{Gyr}$, the kinetic energy injection feedback mode turns on in the simulation for the sole galaxy in our sample that is the main progenitor of its $z=0$ descendant, and at $1.28 \text{Gyr}$ for the remaining galaxies in the sample.

\begin{figure}
	\includegraphics[width=\columnwidth]{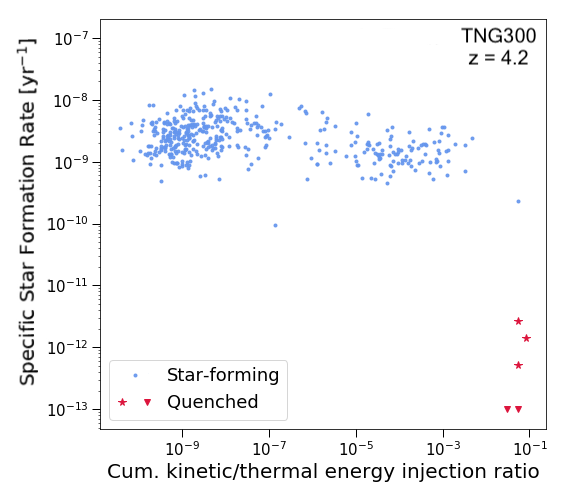}
    \caption{
     sSFR is plotted against the ratio between the cumulative energy injected by the kinetic and thermal AGN feedback modes to show the role of AGN feedback in setting the SFR of galaxies at $z\sim 4.2$ in TNG300. This includes our quiescent sample, along with all star-forming galaxies with $\mathrm{M}_* > 10^9 \mathrm{M}_\odot$. Galaxies with sSFR$=0$ have been brought up to the band at sSFR $=10^{-13}$ [yr$^{-1}$] in downward carets.}
    \label{fig:feedback}
\end{figure}

To account for the potential role of stellar feedback in the quenching processes of our sample, we utilise feedback energy equations provided by \citet{2018MNRAS.479.4056W}. We find that though thermal AGN and stellar feedback dominate at higher redshifts, these rates of feedback energy experience minimal growth during our sample's quenching epoch. Instead, it appears that quenching is largely driven by kinetic AGN feedback, which acts later to efficiently expel remaining reserves of star-forming gas. This corroborates \citet{2018MNRAS.479.4056W}'s finding that kinetic AGN feedback tends to take over at late times in massive haloes in order to keep the star formation rate low.
Fig.~\ref{fig:feedback} displays that the onset of quenching at $z\sim 4.2$ in TNG300 takes place when kinetic mode energy exceeds 1 per cent of the thermal mode.
The necessity of low-accretion state kinetic AGN feedback in the quenching processes of these galaxies is further substantiated by \citet{2017MNRAS.465.3291W, 2018MNRAS.475..624N, 2020MNRAS.493.1888T, 2020MNRAS.499..768Z, 2021MNRAS.508..219N}, who show that TNG galaxies quench only when the kinetic wind mode of feedback is turned on. This mode is more efficient in halting star formation in a host galaxy than the quasar-heating mode, as thermal energy produced by the latter is quickly radiated away.

We conclude that AGN feedback has substantially contributed to the quenching process of the galaxies in our sample. For each galaxy, rapid SMBH growth and activation of a kinetic feedback mode coincide with a steep decline in star formation rate, as evidenced by Fig.~\ref{fig:5SFH}. 
Even so, the cause of stellar feedback’s reduced efficiency in this study remains ambiguous; this feedback mode may play a more pivotal role in the quenching of galaxies in the real Universe. Further observational comparisons of high-redshift quiescent sources are necessary in order to precisely determine the roles of AGN and stellar feedback mechanisms in quenching.

\section{Final conclusions and applications to James Webb}
\label{jwst}

We have studied the attributes and evolution of the first massive quiescent galaxies in the TNG300 simulation of the IllustrisTNG project.
These galaxies' physical characteristics corroborate those of early massive quiescent galaxies directly observed in our Universe, namely a compact stellar half-mass radius and substantial stellar-to-gas mass ratio. 
However, our sample emphasises a critical evolutionary caveat: 
observed high-redshift quenched populations are far from the final word in
the time evolution of the earliest quiescent galaxies. The truncation of star formation that is driven by AGN feedback in IllustrisTNG is only a pause in these developmental paths, which include significant growth and star formation in other branches of their evolutionary trees.
While expanding our stellar mass and sSFR cuts to $10^{10} \mathrm{M}_\odot$ and $10^{-10}$ yr$^{-1}$ did not yield additional TNG quiescent candidates, we intend to fully investigate quenching at lower mass regimes in future research.

With the results of this study, we predict that the first quiescent galaxies to be located by \emph{JWST} will host massive black holes, 
and furthermore that the \emph{JWST} mission will allow for spectroscopic confirmation of the presence of AGN at high redshifts ($4 < z < 5$).
Additional noteworthy \emph{JWST} predictions facilitated by TNG include the works of \citet{2020MNRAS.495.4747S} and \citet{2022arXiv220800007C}. 
\citet{2020MNRAS.495.4747S} provide predictions for the dust attenuation curves of high redshift galaxy populations ($z = 2 - 6$) based on the IllustrisTNG simulation suite, and the authors conclude that attenuation curves are steeper in galaxies at higher redshifts, with bluer colours, or with lower stellar masses. 
\citet{2022arXiv220800007C} present a catalogue of mock images of massive high-redshift galaxies from the TNG50 cosmological simulation. The authors analyse the predictions of TNG50 for the size evolution of galaxies at $3 \le z \le 6$ and the expectations for CEERS to probe that progression, finding a difference between the mass and light distribution, which may indicate a transition in the galaxy morphology at $z = 4 - 5$. \emph{JWST} findings shall further test these predictions.


Furthermore, \emph{JWST} has facilitated the selection of a sample of massive quiescent galaxies at $3 < z < 5$ by \citet{2022arXiv220800986C}.
The number density reported by the authors 
for massive quiescent galaxies at $4 < z< 5$, $n = 2.3^{+3.1}_{−1.5} \times 10^{−5} \text{Mpc}^{−3}$, is
in agreement with the calculated number density of our sample, $n = 1.3 \times 10^{−5} \text{Mpc}^{−3}$.
This observational number density represents a conservative lower limit, indicating that \emph{JWST} may reveal new insights into galaxy formation that will train the next generation of cosmological simulations. 
Our study serves as a strong test of the quenching mechanisms employed in such simulations; if larger samples of quiescent galaxies are observationally confirmed at higher redshifts than 
$z\sim4.2$, as suggested by photometry \citep[e.g.,][]{2019MNRAS.490.3309M, 2020ApJ...897...44S}, it implies an interesting discrepancy to pursue with further research of high-redshift galaxies.

\section*{Acknowledgements}

We are immensely appreciative of the  IllustrisTNG team for making their simulation public \citep{2018MNRAS.475..624N}. 
This research used the Python packages Numpy \citep{2020Natur.585..357H} and  Matplotlib \citep{2007CSE.....9...90H}.
A.I.H. gratefully acknowledges the financial support of a CU Boulder Undergraduate Research Opportunities Program Individual Grant.


\section*{Data Availability}

The data utilized in this paper were accessed from the IllustrisTNG simulations, which are publicly available at \url{www.tng-project.org/data}. This includes the TNG300 simulation used in this study.



\bibliographystyle{mnras}
\bibliography{example} 








\bsp	
\label{lastpage}
\end{document}